\documentclass[12pt]{article}
\usepackage{makeidx}
\usepackage{multirow}
\usepackage{multicol}
\usepackage[dvipsnames,svgnames,table]{xcolor}
\usepackage{graphicx}
\usepackage{epstopdf}
\usepackage{ulem}
\usepackage{hyperref}
\usepackage{amsmath}
\usepackage{amssymb}
\author{Yuri Kamyshkov}
\title{}
\usepackage[paperwidth=612pt,paperheight=792pt,top=72pt,right=72pt,bottom=72pt,left=72pt]{geometry} 

\def\bef{\begin{figure}}
\def\eef{\end{figure}}

\newcommand{\be}[1]{\begin{equation}\label{#1}}
\newcommand{\beq}{\begin{equation}}
\newcommand{\ee}{\end{equation}}
\newcommand{\beqn}[1]{\begin{eqnarray}\label{#1}}
\newcommand{\eeqn}{\end{eqnarray}}
\newcommand{\bd}{\begin{displaymath}}
\newcommand{\ed}{\end{displaymath}}

\def\lsim{\raise0.3ex\hbox{$\;<$\kern-0.75em\raise-1.1ex
e\hbox{$\sim\;$}}}
\def\gsim{\raise0.3ex\hbox{$\;>$\kern-0.75em\raise-1.1ex
\hbox{$\sim\;$}}}
\def\simlt{\mathrel{\lower2.5pt\vbox{\lineskip=0pt\baselineskip=0pt
           \hbox{$<$}\hbox{$\sim$}}}}
\def\simgt{\mathrel{\lower2.5pt\vbox{\lineskip=0pt\baselineskip=0pt
           \hbox{$>$}\hbox{$\sim$}}}}
\def\unity{{\hbox{1\kern-.8mm l}}}
\newcommand{\ov}{\overline}
\renewcommand{\to}{\rightarrow}

\newcommand{\dm}{\varepsilon}

\newcommand{\nbar}{{\bar n} }
\newcommand\PM[1]{\begin{pmatrix}#1\end{pmatrix}}

\renewcommand{\to}{\rightarrow}

\def\lsim{\mathrel{\mathop  {\hbox{\lower0.5ex\hbox{$\sim$}
\kern-0.8em\lower-0.7ex\hbox{$<$}}}}}
\def\gsim{\mathrel{\mathop  {\hbox{\lower0.5ex\hbox{$\sim$}
\kern-0.8em\lower-0.7ex\hbox{$>$}}}}}
\def\cN{{\cal N}}
\def\cM{{\cal M}}
\def\cO{{\cal O}}
\def\cO{{\cal O}}

\makeatletter
	{\par\setlength{\parindent}{#3}
	\setlength{\leftmargin}{#1}       \setlength{\rightmargin}{#2}%
	\advance\linewidth -\leftmargin       \advance\linewidth -\rightmargin%
	\advance\@totalleftmargin\leftmargin  \@setpar{{\@@par}}%
	\parshape 1\@totalleftmargin \linewidth\ignorespaces}{\par}%
\makeatother 

\begin{document}

\begin{center}
\textbf{{\large Gauged $B-L$ Number and  Neutron--Antineutron Oscillation: \\
\vspace{2mm}
Long-range Forces Mediated by Baryophotons }}
\end{center}

\begin{center}
\textit{Andrea Addazi$^{1}$, Zurab Berezhiani$^{2}$}
\end{center}

%\vspace{-2mm} 

\begin{center}
\textit{{\small Dipartimento di Fisica e Chimica, Universit\'a di L'Aquila, 67010 Coppito, L'Aquila (AQ) 
and INFN,  Laboratori Nazionali del Gran Sasso, 67010
Assergi (AQ), Italy}}
\end{center}

\begin{center}
\textit{Yuri Kamyshkov$^{3}$}
\end{center}

\begin{center}
\textit{{\small Department of Physics, University of Tennessee, Knoxville, TN
37996-1200, USA}}
\end{center}

\begin{abstract}
\small
Transformation of neutron to antineutron is a small effect that has not yet been experimentally observed.  
%\cite{Phillips:2014fgb}. 
In principle, it can occur with free neutrons in the vacuum or with bound neutrons inside the nuclear environment different for neutrons and antineutrons and for that reason in the latter case it is heavily suppressed. Free neutron transformation also can be suppressed if environmental vector field exists distinguishing neutron from antineutron. 
We consider here the case of a vector field coupled to $B-L$ charge of the particles ($B-L$ photons) 
and study a possibility of this to lead to the observable suppression of neutron to antineutron transformation. 
The suppression effect however can be removed by applying external magnetic field.
If the neutron--antineutron oscillation will be discovered in  free neutron oscillation experiments, this will imply 
limits on $B-L$ photon coupling constant and interaction radius 
few order of magnitudes stronger than present limits form the tests of the equivalence principle. 
If $n-\nbar$ oscillation will be discovered via nuclear instability, but not in free neutron oscillations in 
corresponding level, this would indicate to the presence of fifth-forces mediated by such baryophotons. 
\end{abstract}

%\end{document}
\section{Introduction}

The oscillation phenomenon between the neutron and antineutron, 
%This mixing induces a very interesting phenomenon of neutron--antineutron oscillation, 
$n \to  \nbar$  was suggested in early 70's  by Kuzmin \cite{Kuzmin:1970nx}. 
First theoretical scheme for $n-\nbar$ oscillation was suggested in Ref. 
\cite{Mohapatra:1980qe}, 
%\cite{Glashow,Mohapatra:1980qe}. 
followed by other models as e.g. \cite{Babu:2001qr,Berezhiani:2005hv,Babu}. 
Experimental observation of the transformation of neutron to antineutron 
$n\to \nbar$ would be a demonstration of baryon number violation by two units, 
from $B=+1$ for neutron to $B=-1$ for antineutron. 
This will be an experimental demonstration that one of the Sakharov's conditions  \cite{Sakharov} 
required for the generation of baryon asymmetry in the universe is indeed realized in the nature.  
The $n\to \nbar$ conversion so far was not experimentally observed. 
However, this does not exclude the possibility 
that it can be a rare/suppressed process.  
For a review of the present theoretical and experimental  situation 
on $n-\nbar$ oscillation, see \cite{Phillips:2014fgb}.  

Apart of baryon-conserving (large) Dirac mass term $m_n \ov{n} n$,
 the neutron may acquire a  small Majorana mass term,  
$\dm_{n\nbar} (n C n   + {\rm h.c.})$,  which violates the baryon number  by two units 
and induces the neutron -- antineutron mass mixing.  
As far as the neutron is a composite particle, $n-\nbar$  mixing can be induced by the 
effective six-fermion operators involving the first family quarks $u$ and $d$: 
% which in terms of the Standard model fragments  $u=u_R$, $d=d_R$ and $q = (u, d)_L$  read as
%\footnote{ }
 %
 \be{nn}
 \cO_9  = \frac{1}{\cM^5} (udd udd )
% \big (udd udd \,  + \,  udd qqd \, + \, qqd qqd \big)     ~~~ (B=2)
%\cO_9  = \frac{1}{\cM^5} udd udd   +  \frac{1}{\cM^5} udd qqd  +    \frac{1}{\cM^5} qqd qqd     ~~~ (B=2)
%\frac{b_1}{\cM^5} udd udd \,  + \,  \frac{b_2}{\cM^5} qqd qqd + \dots,   ~~~ (\Delta B=2)
% + {\rm h.c.}  \;.
\ee
 where $\cM$ is some large mass scale of new physics beyond the Standard Model.  
 These operators can have different convolutions of the Lorentz, color and weak isospin indices 
 which are not specified.  
% (Needless to say, the combination $qq$ in second term in (\ref{S}) must be in a weak isosinglet combination, 
%$qq =\frac12 \epsilon^{\alpha\beta} q_{\alpha}q_{\beta} = u_L d_L $ where 
%$\alpha,\beta= 1,2$ are the weak $SU(2)$ indices, while in the third term $qq$ can be taken
% in a weak isotriplet combination as well.)   
% and $b_{1,2}$ are the order one dimensionless constants.  
%of the order of one. 
%These operators can have different Lorentz and color  structures  
%\footnote{
(More generally, 
having in mind  that all quark families can be involved, 
such operators can induce the mixing phenomena  also  for other neutral baryons, e.g.  
between the hyperon $\Lambda$ into the anti-hyperon $\bar\Lambda$.)
The models of Refs. \cite{Mohapatra:1980qe,Babu:2001qr,Berezhiani:2005hv,Babu} are just 
different field-theoretical realizations for the operators (\ref{nn}).

Taking matrix elements from these operators between the neutron and antineutron states, 
one can estimate the neutron Majorana mass modulo the Clebsch factors as   
\be{dm}
\dm_{n\nbar} \sim  \frac{\Lambda_{\rm QCD}^6}{\cM^5} \sim   \left(\frac{1 \, {\rm PeV} }{\cM}\right)^5 \times 
10^{-25} \, {\rm eV}	\; .
\ee
The  coefficients of matrix elements $\langle \nbar \vert uddudd  \vert n\rangle$ 
for different Lorentz and color structures of operators (\ref{nn}) were studied in ref. \cite{Rao},  
but we do not concentrate here on these particularities and take them as $O(1)$ factors.  

Concerning the experimental limits on $n-\nbar$ oscillation time $\tau_{n\nbar} = 1/\dm_{n\nbar}$,  
the direct limit on free neutron oscillations imply $\tau_{n\nbar} > 0.86 \times 10^8$~s \cite{Grenoble}. 
The nuclear stability limits, with uncertainties  in the evaluation of nuclear matrix elements, 
translate into  $\tau_{n\nbar} > 1.3 \times 10^8$~s \cite{Soudan}
and $\tau_{n\nbar} > 2.7 \times 10^8$~s \cite{SK2015}. The latter implies the strongest upper limit 
on $n-\nbar$ mixing,  $\dm_{n\nbar} < 2.5 \times 10^{-24}$~eV. 
The future long-baseline direct experiment at the European Spallation Source (ESS) can 
reach the sensitivity down to $10^{-25}$~eV and thus improve the existing limits 
on $n-\nbar$ oscillation time by more than 
an order of magnitude \cite{Phillips:2014fgb}.

One can consider a situation when baryon number $B$ is broken not explicitly but spontaneously.  
Such a baryon symmetry can be global or local, with different physical implications. 

 The possibility of spontaneous violation of global lepton symmetry 
 after which the neutrinos can get non-zero Majorana masses   is widely discussed  in the literature. 
%  This could happen if a scalar $\chi$  with $L=-2$ gets a non-zero VEV $\langle \chi \rangle = V$, 
  As a result, a Goldstone boson should appear in the particle spectrum, named as 
 Majoron \cite{Chikashige:1980ui}. 
% Its coupling constant to a neutrino  $\nu$ is related to the neutrino mass $m_\nu$ 
% as $g_\nu = m_\nu/V $.  
 Spontaneous violation of global baryon number in connection with the Majorana 
 mass of the neutron
 %, i.e. neutron antineutron oscillation, 
 was first discussed in ref. \cite{Barbieri}, 
 in the context of Mohapatra-Marshak model for $n-\nbar$ oscillation \cite{Mohapatra:1980qe}. 
% However this concept did not attract much attention in the literature. 
Recently the discussion was revived by one of us in Ref. \cite{baryo-majoron}, where also seesaw 
 model for $n-\nbar$ transition with  low scale spontaneous violation of baryon number 
 were suggested.  
 An associated Goldstone particle -- baryo-majoron, can have observable 
effects in neutron to antineutron transitions in nuclei or dense nuclear matter.  
The low-scale baryo-majoron model \cite{baryo-majoron} has many analogies with the 
low scale Majoron model for the neutrino  masses  \cite{Berezhiani:1992cd}. 
By extending baryon number  to $B-L$ symmetry, baryo-majoron can be identified with 
the ordinary majoron associated with the spontaneous breaking of lepton number, 
with interesting implications for neutrinoless $2\beta$ decay with the majoron emission \cite{Georgi:1981pg}, 
for matter-induced effects of the neutrino decay  \cite{Berezhiani:1987gf} and  
for the Majoron field effects in the early Universe \cite{Bento:2001xi}. 
 
In this paper we discuss a situation when baryon number is related to a local gauge symmetry. 
The idea  to describe the conservation of baryon number $B$ and lepton number $L$ 
similar to the conservation of electric charge by introducing gauge symmetries $U(1)_B$ and $U(1)_L$,  
i.e. in terms of baryon or lepton charges coupled to the massless vector fields 
of leptonic or baryonic photons with tiny coupling constants, was suggested long time ago \cite{OkunL&Q}. 
Their effects for the neutron oscillations were studied in Refs. \cite{Lamoreaux}. 
%Conservation of $B$ and $L$ would result of gauge invariance of massless field $b_{\mu}$.
%Such possibility was discussed in the literature, e.g. 
Nowadays the limits on such interactions are very stringent.  Best limits on the coupling strength of 
baryonic and leptonic photons were obtained from the E\"{o}tv\"{o}s type of experiments
 testing the equivalence principle~\cite{Adelberger}. 
Then, the common sense argument used here was that coupling of such photons is many order of magnitude weaker 
than the gravitational interaction between baryons or leptons and therefore such photons are likely non-existent.
We will try to revise this concept. 

Since baryon number $B$ and  lepton number $L$ separately are not conserved  
due to non-perturbative effects, it is difficult to promote them as gauge symmetries without 
altering the particle content of the Standard Model, i.e. without introducing new exotic particles. 
%let us consider a combined quantum number $Q=B-L$.  
Therefore, we  discuss not baryonic and leptonic photons separately  but the vector fields 
associated with $U(1)_{B-L}$ gauge symmetry. 
%$U(1)_B$ and $U(1)_L$ separately, 
In the Standard Model this symmetry is anomaly free  and $B-L$ current 
is conserved at the perturbative as well as at non-perturbative level. On the other side, 
the existence of neutron--antineutron oscillation or other similar phenomena 
 would imply that this gauge symmetry, if it exists, should be  spontaneously broken. 
 $n-\nbar$ mixing cannot be induced without violating $B$ and thus $B-L$, which should 
render massive  also $B-L$ baryophoton. 
  %and thus corresponding baryophotons should become massive. 
However, if its gauge coupling constant is very small,  such a baryophoton 
can remain extremely light, 
%with masses $\sim 10^{-17}$ eV or so, 
and mediate observable long range forces (fifth force) between material bodies. 
%provided that gauge coupling constant of baryophoton is tiny enough. 

Clearly, such $B-L$ baryophotons couple with opposite charges not only between 
the baryons and anti-baryons (and between the leptons and anti-leptons), 
but also between the baryons and leptons.
Thus, $B-L$ charge of the neutral hydrogen atom is zero 
while the $B-L$ charge of heavier neutral atoms is determined by the number of neutrons in nuclei. 
Therefore, the regular matter built by nuclei heavier than hydrogen is $B-L$ charged. 
In principle, at the scale of the universe 
$B-L$ charge might be compensated by the relic neutrino component that so far 
remains experimentally undetected. 

%\end{document} 
\section{Experimental limits on B-L photons}
 
The  baryophoton $b_\mu$ associated with $U(1)_{B-L}$ gauge symmetry 
interacts with the fermion (neutron, proton, electron and neutrino) currents as 
$g b_\mu (\ov{n} \gamma^\mu n + \ov{p} \gamma^\mu p - \ov{e} \gamma^\mu e - \ov{\nu} \gamma^\mu \nu)$. 
As far as the existence of the neutron--antineutron mixing implies the violation of $U(1)_{B-L}$, this gauge 
boson cannot remain exactly massless. 

In particular, since  $D=9$ effective operators (\ref{nn}) are now forbidden by $U(1)_{B-L}$ symmetry, 
they can be  replaced by the effective $D=10$ operators
 \be{chi}
  \frac{\chi}{M^6}(udd udd)
\ee  
%  ($\Delta B = 0$), 
involving the complex scalar  field $\chi$ bearing two units of $B-L$ number, $Q_\chi=-2$.  
%Hence, the $n-\nbar$ mixing mass 
Its vacuum expectation value (VEV) $\langle \chi \rangle = \upsilon_\chi$.  
spontaneously breaks the $U(1)_{B-L}$. By substituting   $\chi \to \langle \chi \rangle$ in (\ref{nn}), 
operator (\ref{chi}) reduces to (\ref{nn}) with $\cM^5 = M^6/\upsilon_\chi$, and thus 
the induced $n-\nbar$ mixing mass 
%is induced via the VEV $\langle \chi \rangle = \upsilon_\chi/\sqrt2$, which 
can be estimated as 
\be{dm-chi}  
 \dm_{n\nbar} \sim  \frac{\upsilon_\chi \Lambda_{\rm QCD}^6}{M^6} \sim   
 \left(\frac{\upsilon_\chi}{1 \, {\rm keV} }\right) \left(\frac{10 \, {\rm TeV} }{M}\right)^6  \times 10^{-25} \, {\rm eV}	\; .
\ee
Therefore, taking that the scale $M$ larger than 10 TeV, the neutron--antineutron oscillation can be 
within the experimental reach at the ESS, i.e. $ \dm_{n\nbar} > 10^{-25}$~eV, if $\upsilon_\chi > 1$~keV or so. 
We take this as a benchmark value for the $B-L$ symmetry breaking scale. 
If the scale $M$ is taken ad extremis as small as $M\sim 1$~TeV, then one would get $\upsilon_\chi \sim 1$~meV. 
However, such a tiny scale does not seem to be a really realistic,  
However, a huge hierarchy problem between the scale $\upsilon_\chi \sim 1$~meV 
and the electroweak scale $\sim 100$ GeV will be a headache. More important, it is very very unlikely 
that the violation of $B-L$ at such a small $\upsilon$, which is just about a 3 K 
(cosmic microwave background (CMB) temperature today), can be relevant for primordial baryogenesis. 
The possibility of very small $\upsilon$ is excluded in realistic models which 
we discuss later.\footnote{The lower limit on the $U(1)_{B-L}$ 
breaking scale becomes  even more stringent if instead of $D=10$ operators one introduces $D=11$ 
operator $\frac{\eta^2}{M^7}(udd udd)$ involving a scalar $\eta$ with the charge $Q=-1$. 
Then for achieving e.g. $ \dm_{n\nbar} \sim 10^{-24}$~eV, it should have larger VEV, 
 $\upsilon_\eta \sim 10$~keV, even if $M\sim 1$~TeV. } 

 In principle, also the VEVs $v_i$ of other scalars $\eta_i$ 
with non-zero $B-L$ charges $Q_i$ can also participate in breaking $U(1)_{B-L}$. 
As a result, 
%of spontaneous $B-L$ breaking by the VEV of scalar $\chi$, 
the baryophoton should acquire the mass:  
 \begin{equation}\label{eq:7}
M_b=  2 \sqrt 2 g \upsilon, \quad \quad  
\upsilon = \big[ \upsilon_\chi^2 + (Q_i/Q_\chi)^2 v_i^2 \big]^{1/2} \geq \upsilon_\chi 
%Y_{\chi} \upsilon_\chi = 2 g \upsilon_\chi, 
\end{equation}
where $g$ is gauge coupling constant of baryophotons, 
%$\upsilon_\chi$ is vacuum expectation value (VEV) of scalar $\chi$ bearing two units of 
%in the range 1 MeV - 1 PeV, and $Y_{\chi}=2$ 
%$Y=B-L$ charge. 
Thus, the value of $\upsilon_\chi$ defines the minimal possible value of $M_b$ for a given constant $g$. 
If there are other scalar fields $\eta_i$ with non-zero 
$Q=B-L$ charges and non-zero VEVs, their contribution would make $M_b$ larger. 
% violating  number (via baryons) by 2 units.

Therefore, baryophotons  should mediate an Yukawa-like fifth-force between 
the material bodies. Vector boson $b_\mu$ exchange induces a spin-independent 
potential energy of the interaction between the test particle 
with $B-L$ charge $Y_i$, in our case the  neutron or antineutron, 
%($Y_n=1$)  and antineutron ($Y_\nbar=-1$), 
and an attractor (a massive body as e.g. Earth or sun) 
with the overall $B-L$ charge $Y_A$: 
%Namely, any massive body as e.g. Earth and sun should produce,  
%besides the gravitational (Newtonian) potential, a ``fifth" Yukawa like potential:  
\begin{equation}\label{eq:6}
V_i  = \alpha_{B-L}\frac{Q_i Q_A}{r}\, e^{-r/\lambda}, \quad \quad \quad 
\lambda = \frac{1}{M_b} 
\simeq \left(\frac{10^{-49}}{\alpha_{B-L} }\right)^{1/2} 
\left(\frac{1~\rm keV}{\upsilon }\right) \times 0.6 \cdot 10^{16} ~ {\rm cm} 
%\vspace{2 mm}
\end{equation}
where $\alpha_{B-L} = g^2/4\pi$,  
in addition to the gravitational potential energy $V_i^{\rm gr} = - G m_i M_A/r$, 
$G$ being the Newton constant.  
The overall $B-L$ charge of the gravitating body of mass $M_A$ 
is defined by its chemical composition:   
$Q_A \simeq Y_n M_A/m_n$,  $m_n$ being the neutron mass.  
Due to electric neutrality, the amount of protons and electrons should be equal and 
thus their contributions cancel each other. Hence, the value of $Q_A$ 
is determined  by the neutron fraction $Y_n$. In particular, $Q=0$ for hydrogen 
and $Q\approx 0.5$ for a typical heavy nuclei.   
%$V_{n}$ and $V_{\nbar}$  of different sign: 
%$V_{n,\nbar} = Y_{n,\nbar} \Phi_b = \pm \Phi_b$. 

%$\lambda = 1/M_b $ is the radius of the $B-L$ Yukawa force, 
The maximal possible range of the Yukawa radius $\lambda$ for a given constant $g$ 
is limited by the minimal value of the symmetry breaking scale, $\upsilon = \upsilon_\chi$. 
If there are other scalar fields $\eta_i$ with different $Q=B-L$ charges and non-zero VEVs, 
their contribution would make the mass $M_b$ larger, and thus would shorten the range of $\lambda$.  
The results of torsion-balance tests of the weak equivalence principle from Ref.~\cite{Adelberger} 
can be interpreted as limits on fifth forces, and in particular, for the force mediated  by the 
$B-L$ baryophotons,  as limits on dimensionless constant $\alpha_{B-L}$ 
for a given radius $\lambda$, as shown in Fig. 1. 

In difference from universal gravity, the baryophoton exchange would induce for the neutron ($Q_n=1$) 
and antineutron ($Q_{\nbar}=-1$) the potential energies of different sign, $V_{\nbar} = - V_{n}$.   
It is convenient to relate the values $V_{n,\nbar}$ with the neutron (and antineutron) gravitational 
potential energy $V_{\rm gr} = -G m_n M_A/r$: 
\be{adelb}
V_{n,\nbar} = \pm \tilde\alpha q_A  e^{-r/\lambda}  \times V_n^{\rm gr} 
\ee
%$\alpha_{B-L}$ to the gravitational (Newton) constant $G$ 
introducing  a dimensionless parameter  $\tilde\alpha = \alpha_{B-L}/G m_n^2$, 
% $Gm_n^2 = 0.59 \times 10^{-38}$   $G$ being the Newton constant.    
and $q_{A} = Q_A/(M_A/m_n)$ being the massive objects $B-L$ charge per neutron mass unit. 
The upper limits on the parameter  $\tilde\alpha$ 
as a function of the radius $\lambda$ are given in Fig. 6 of Ref. \cite{Adelberger} 
(in Ref. \cite{Adelberger} these values are normalized per 
atomic mass unit, 1~amu$ = 0.99\, m_n$). 
In Fig. 1 these limits  of Ref. \cite{Adelberger} are shown directly translated for $\alpha_{B-L}$. 
%for a given radius $\lambda$. 
As we see, if the Yukawa radius is larger than the Earth diameter, 
$\lambda > 10^{9}$~cm or so, 
then the upper limit on $\alpha_{B-L}$ becomes practically independent on $\lambda$  
and it corresponds to $\alpha_{B-L} < 10^{-49}$, 
or $\tilde\alpha < 1.7\times 10^{-11}$ \cite{Adelberger}.

\begin{figure}[t]	
	\centering
%%%%%%%                     trim = l    b    r   t
\includegraphics[width=400pt, trim=40 340 40 60, clip=true]{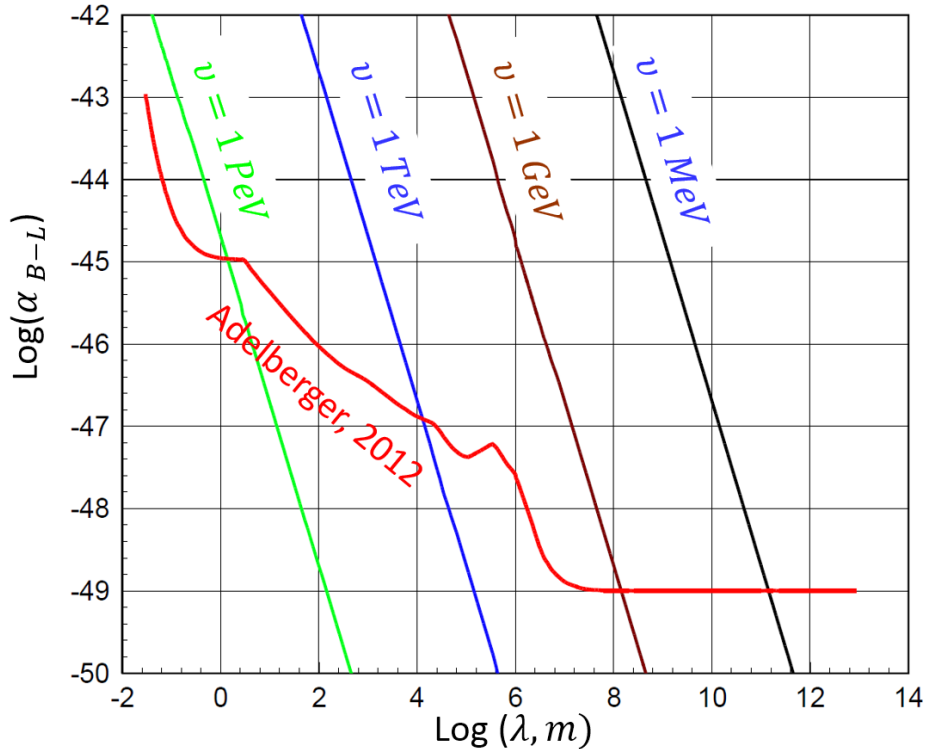}
	\caption{Limits on $B-L$ dimensionless interaction constant (see text). 
	Different values of the VEV $\upsilon_\chi$ 
	responsible for generating of the baryophoton mass are shown here 
for the sake of demonstration of the possible scale of the mechanism. }\label{fig:1}
\end{figure}

Now we can estimate the neutron potential energy $V_n$ as produced 
due to the baryophoton potentials by the Earth, sun and the Galaxy, relative to the 
corresponding gravitational potential energies.   

The Earth induces the gravitational potential energy for the neutron at its surface 
$V^{\rm gr}_E = m_n \phi_{\rm gr}= -Gm_n M_\oplus/R_\oplus \approx 0.66$ eV. 
The sun gives a bigger contribution, $V^{\rm gr}_S = -Gm_n M_\odot/{\rm AU} \approx 10$ eV. 
 Finally, the Galaxy itself induces even bigger value  
 $V^{\rm gr}_G \sim 1$ keV.\footnote{ Notice that we are dealing with 
the gravitational potentials which fall as $\propto 1/r$ and not  with gravitational forces 
testable by torsion balance experiments. 
The latter are $\propto 1/r^2$ and their hierarchy between the Earth, sun and the Galaxy 
becomes reordered in opposite way. This is the reason why for $\lambda$ exceeding the 
Earth Diameter, the experimental limits of Ref. 
\cite{Adelberger} become independent on $\lambda$. } 
%However, the gravitational potential energies 
%for the neutron and antineutron are equal and they should not influence the $n-\nbar$ oscillation. 

Since the Earth is built by heavy nuclei,  
$B-L$ charge of the Earth is approximated as 50\% of the number of baryons in the Earth, i.e. 
$q_E \simeq 0.5$. 
The sun is dominantly consists of hydrogen which has vanishing $B-L$, and thus its fifth force is 
essentially determined by the mass fraction of heavier nuclei (helium, etc.) with $Y_n \simeq 0.5$. 
Therefore, $q_{S}   \simeq 0.13$ as one can estimate  from the known 
chemical composition of Sun~\cite{Suncomp}. The same applies to the Milky Way contribution, 
$q_G \simeq 0.13$. 

Thus, assuming that $\lambda$ is larger than the Earth diameter, $\lambda > 2R_\oplus$, 
the values of $V_n$ at the surface of Earth can be estimated as:\footnote{We neglect 
the annual modulation of $V^{\rm gr}_S$ due to small variation 
the sun--Earth distance, as well as potentials induced by  other planets and the Moon. 
The latter also could be responsible for time variation of the total potential. We also neglect 
contributions from neighboring galaxies and galaxy clusters since 
}
\begin{equation}\label{eq:8}
{V_n} = \tilde \alpha  
\times \big( 0.5 \, V_E^{\rm gr} e^{-R_\oplus/\lambda} +  0.13\, V_S^{\rm gr}  e^{-1~{\rm AU}/\lambda} +  
0.13\, V_G^{\rm gr} e^{-10~{\rm kpc}/\lambda}  \big ) 
\end{equation}
From (\ref{eq:6}), taking $\alpha_{B-L} = 10^{-49}$, i.e. $\tilde\alpha = 1.7 \times 10^{-11}$, 
we see that for our benchmark value 
$\upsilon_\chi = 1$~keV we obtain $\lambda \sim 10^{16}$~cm which is much larger than the 
sun-Earth distance (1 AU$ \approx 1.5 \times 10^{13}$~cm). Therefore, in the case 
the  contributions of the Earth and the sun in $V_n$  can be as large 
as respectively $0.56\times 10^{-11}$~eV and $2.2\times 10^{-11}$~eV,   
amounting in total as $2.8 \times 10^{-11}$~eV. For $\lambda$'s smaller than 
the Earth diameter,  the larger values of $\alpha_{B-L}$ are allowed (see Fig. 1) but 
the available volume of the source drops as $(\lambda/R_\oplus)^3$ and thus 
upper limit on $V_n$ sharply decreases.  

It is interesting to question to how large values of $\lambda$ and  how large potential $V_n$ 
can be induced by the Galaxy. For the benchmark value $\upsilon_\chi = 1$~keV, we 
see from (\ref{eq:6}) 
that for having $\lambda > 20~{\rm kpc} = 6\times 10^{22}$~cm, one has to take 
$\alpha_{B-L} < 10^{-62}$ or so, in which case the baryophoton induced potential 
will be less than $10^{-21}$~eV and therefore it would have no influence 
for the experimental search of $n-\nbar$ oscillations. 
Even taking the value of the VEV 
 as small as $\upsilon_\chi = 1$~meV, i.e. 
 at its extreme dictated by the value $\dm_{n\nbar} \sim 10^{-24}$~eV by 
 the operators (\ref{chi}) with $M = 1$~TeV, 
we obtain that $\lambda > 10~{\rm kpc} = 3\times 10^{22}$~cm can be obtained  
if $\alpha_{B-L} < 6\times 10^{-51}$ or so. In this marginal situation, 
the Galaxy contribution in $V_n$ could amount up to  $10^{-10}$~eV.
%a value up two an order of magnitude  larger than the solar contribution, and  
In any case, contribution of more distant objects as neighboring galaxies and galaxy clusters 
are exponentially suppressed since a very small $B-L$ breaking scale, $\upsilon_\chi < 1$~meV, 
is not of interest. 
%It would imply $\dm_{n\nbar} < 10^{-25}$~eV, i.e. neutron--antineutron oscillation 
%principally out of experimental sensitivity  for next 100 years. 

\section{$n-\nbar$ oscillation in the presence of $B\!-\!L$ fifth force }

Non-relativistic Hamiltonian that describes $n-\nbar$  oscillation in the presence of 
fifth force and magnetic fields can be presented 
as $4\times 4$ matrix acting on the state vector $(n_+,n_-,\nbar_+, \nbar_-)$ describing the neutron 
and antineutron states with two spin polarizations:\footnote{Let us recall that the CPT invariance 
implies that the neutron and antineutron must have exactly equal masses and magnetic moments 
of the opposite sign.  
}
\begin{equation}\label{eq:1}
H = \left(\begin{array}{cc}
m_n(1 - \phi_{\rm gr}) + V_n + \mu_n B \sigma_3 & \dm_{n\nbar} \\
\dm_{n\nbar} & m_n(1 - \phi_{\rm gr}) + V_{\nbar} - \mu_n B \sigma_3 
\end{array}\right) , 
%\vspace{3 mm}
\end{equation}
where $\mu_n = -6 \times 10^{-12}$ eV/G is the magnetic moment of the neutron, 
$B$ is the magnetic field and $\sigma_3$ is the third Pauli matrix since the spin quantization axis 
is chosen as the direction of the magnetic field. In this basis one has no spin precession and 
%In fact, taking the spin quantisation axis towards the magnetic field direction, 
the Hamiltonian (\ref{eq:1}) is diagonal. Omitting the universal terms and taking 
$V = V_n = -V_{\nbar}$, it can be rewritten as
\be{mat44}
H_I = \PM{ V - \Omega_B & 0 & \dm_{n\nbar} & 0 \\
0 & V + \Omega_B & 0 & \dm_{n\nbar}  \\
\dm_{n\nbar}  & 0 & - V + \Omega_B  & 0 \\
0 & \dm_{n\nbar}  & 0 & - V - \Omega_B  }   .
\ee
%where $\Omega_B = \vert \mu_n B \vert = 6\cdot 10^{-12} (B/10^5\,{\rm nT})$~eV is the Zeeman
where $\Omega_B = \vert \mu_n B \vert = 6\cdot 10^{-12} (B/1\,{\rm G})$~eV is the Zeeman 
energy shift induced by the magnetic field. In general case, with $V_n$ and $\Omega_B$ both non-zero, 
the $n$ and $\nbar$ oscillation probabilities are different between  the  $+$ and $-$ polarization states:    
\begin{equation}\label{eq:3}
P^{\pm}_{n\nbar}(t) =\frac{{\dm_{n\nbar}}^2}{\dm_{n\nbar}^2+ \Delta_{\pm}^2}
 \sin^2\left(t \sqrt{\dm_{n\nbar}^2+ \Delta_{\pm}^2 } \right), 
 \quad\quad \Delta_{\pm} = V \mp \Omega_B  
% {\sin}^2\frac{\sqrt{{\epsilon{}}^2+{\Delta{}V}^2}}{\hslash{}}t
\end{equation}
where $t$ is a neutron free flight time. 
In the realistic experimental conditions $t$  cannot be very large, 
e.g. it was $\sim 0.1$~s in the experiment \cite{Grenoble}, it  can be up to  $\sim 1$~s in the 
experimental setup for cold neutrons at the ESS, and in principle it could reach $\sim 10$~s 
in the experiments when the neutrons vertically fall down in a deep mine.  

Let us discuss first the case when the fifth force is absent, $V=0$, and there remains only  
the magnetic field contribution, i.e.  $\Delta_{\pm}^2 = 4\Omega_B^2\gg \dm_{n\nbar}^2$. 
Then the neutron oscillation probabilities of $+$ and $-$ polarization states should be equal 
\begin{equation}\label{eq:3}
P^{\pm}_{n\nbar}(t) = \frac{{\dm_{n\nbar}}^2}{\dm_{n\nbar}^2+ \Omega_{B}^2}
 \sin^2\left(t \sqrt{\dm_{n\nbar}^2+ \Omega_{B}^2 } \right), 
\end{equation}
and for making effective the oscillation during a time $t$, the magnetic field should be suppressed 
to a needed degree. Namely, If $\Omega_B t \gg 1$, the oscillations should be averaged in time and one gets 
$P^{\pm}_{n\nbar} = \dm_{n\nbar}^2/2\Omega_B^2$.  
%$P^{\pm}_{n\nbar} = \frac{\dm}{\Delta}$ 
However, for small free flight times, $t < 1$~s,  the magnetic field can be suppressed achieving $\Omega_B t < 1$. 
 The argument of sine wave is small and oscillation probability $P(t)$ becomes practically 
 independent on $\Omega_B$:  
% (that means that $\Delta V$ remains $\lesssim 10^{-16}$ eV) the probability can be described as
%
\begin{equation}\label{eq:4}
P \approx \left(\dm_{n\nbar} t \right)^2= \left(t/\tau_{n\nbar} \right)^2
\end{equation} 
The latter condition is known as "quasi-free" condition. The needed level of magnetic field suppression 
depends on the neutron free flight time in experimental conditions. For $t=0.1$~s, the condition 
$\Omega_B <  t^{-1}$ implies $\Omega_B < 10^{-15}$~eV, and thus $B < 10^{-4}$~G.  
Suppressing fields to the level of 1 nT would be sufficient for future realistic experimental times order 1 s.

\begin{figure}	
	\centering
%%%%%%%                     trim = l    b    r   t
\includegraphics[width=400pt, trim=40 340 40 100, clip=true]{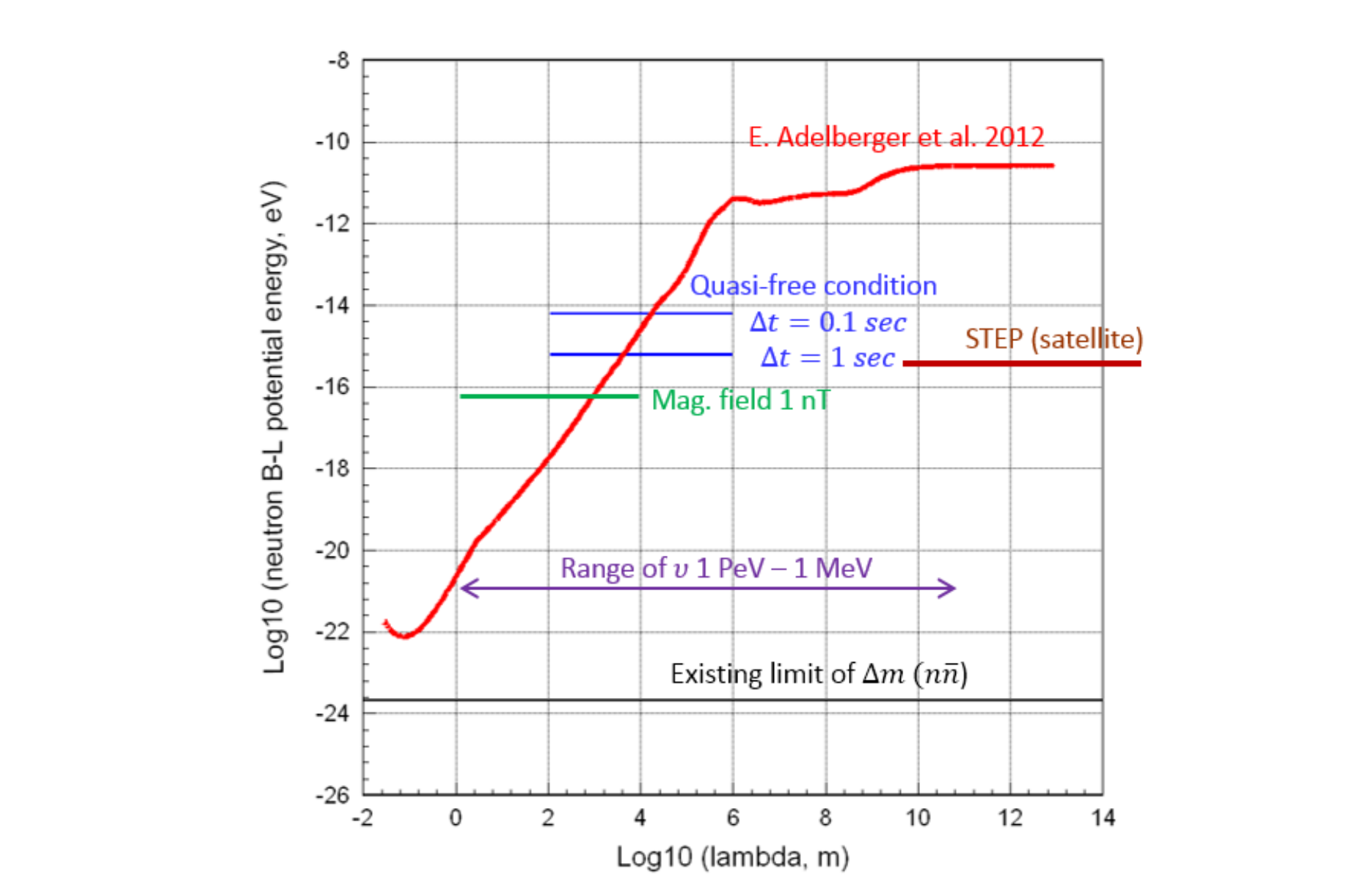}
	\caption{Potential energy $V_n$ of the neutron in $B-L$ field of Sun and Earth. 
	Region of potentils $V_n$ above the red curve is excluded by torsion balance experiment 
	\cite{Adelberger}.}\label{fig:2}
\end{figure}

Let us consider now the case with non-zero $V_n$. 
From (\ref{eq:6}), taking $\alpha_{B-L} = 10^{-49}$, we see that for our benchmark value 
$\upsilon_\chi = 1$~keV we obtain $\lambda \sim 10^{16}$~cm which is much larger than the 
sun-Earth distance (1 AU$ = 1.5 \times 10^{13}$~cm).   
We see from Fig. 1, that for $\lambda > 1$ AU, 
$V_n$ can reach the values up to $3\times 10^{-11}$~eV, equivalent to $\Omega_B$ of the magnetic field 
$B \simeq  5$~G. This would lead to strong suppression of $n-\nbar$ oscillation even if the magnetic 
field value vanishing: the $n-\nbar$ oscillation will not be discovered at the ESS even
 if $\dm_{n\nbar} > 10^{-24}$~eV.  
Therefore, for achieving the quasi free condition for $n-\nbar$ oscillation allowing to discover $n-\nbar$ conversion, 
the value of magnetic field should be tuned with precision of few nT to a resonance value, 
so that $\Omega_B = V_n$ with the precision of $10^{-16}$ eV or so. Let us noticed, that 
since oscillation probabilities of $+$ and $-$ polarization states are different, see eq. (\ref{eq:3}), 
resonance can occur for only for one polarization. 
%The magnetic field dependence of the oscillation 
%probability averaged over the spectrum is shown of Fig. 3. 

Levels of potential energy $V_n$ corresponding to quasi-free conditions for $n \to \nbar$ 
observation time $\Delta{t}=0.1$ and $1.0$ s are shown in the Fig. 2 together with 
$\Omega_B$ corresponding to the  magnetic field 1 nT. 
We see that $V_n$ can exceed the limit of quasi-free condition 
in the range of $\lambda$ between $\sim 10^{4} - 10^{13}$ m. 
In this region $n \to \nbar$ oscillation can be suppressed. 

However, tiny fifth forces have no effect in intranuclear $n \to \nbar$ transformations. 
One can envisage scenario where $n \to \nbar$ will be discovered 
in intranuclear transformations in large underground experiments 
although it will not be observed in transformation with free neutrons at the corresponding level, e.g. at the ESS. 
This can be an indication that some extra potential different between the neutron and antineutron is in play, 
which can be induced by $B-L$ photons under considerations. 
This situation can be checked by applying in free neutron experiments the magnetic field with programmed 
magnitude and direction in the whole neutron flight path and by varying of this field to find the resonance 
value for which it would compensate the effect of $B-L$ field induced potential $V_n$. 
Example of such variation of magnetic field is shown in Fig. 3 assuming that $V_n=10^{-12}$ eV. 

\begin{figure}[t]	
	\centering
%%%%%%%                     trim = l    b    r   t
\includegraphics[width=400pt, trim=40 340 40 60, clip=true]{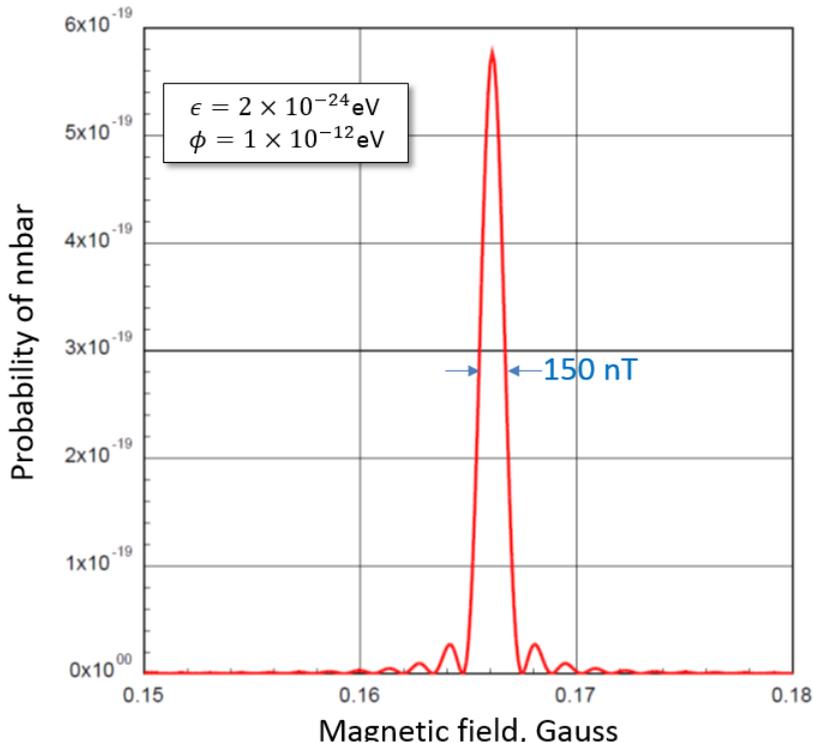}
	\caption{Variation of external magnetric field that would compensate the suppressing effect of the $B-L$  potential (see text)}\label{fig:3}
\end{figure}

\section{Low scale seesaw model} 

Is it possible to built a consistent model in which baryon number, or $B-L$,    
spontaneously breaks at rather low scales in which case the baryophoton couplings 
 to the neutron can have an effect on the laboratory search of $n-\nbar$ oscillation?  
% scale $\sim 1$~MeV or even smaller?  
%The answer is positive. 
%This can be  obtained by a simple modification of the above considered model. 

\begin{figure}[t]
\begin{center}
\vspace{-1.cm}
\includegraphics[width=8cm]{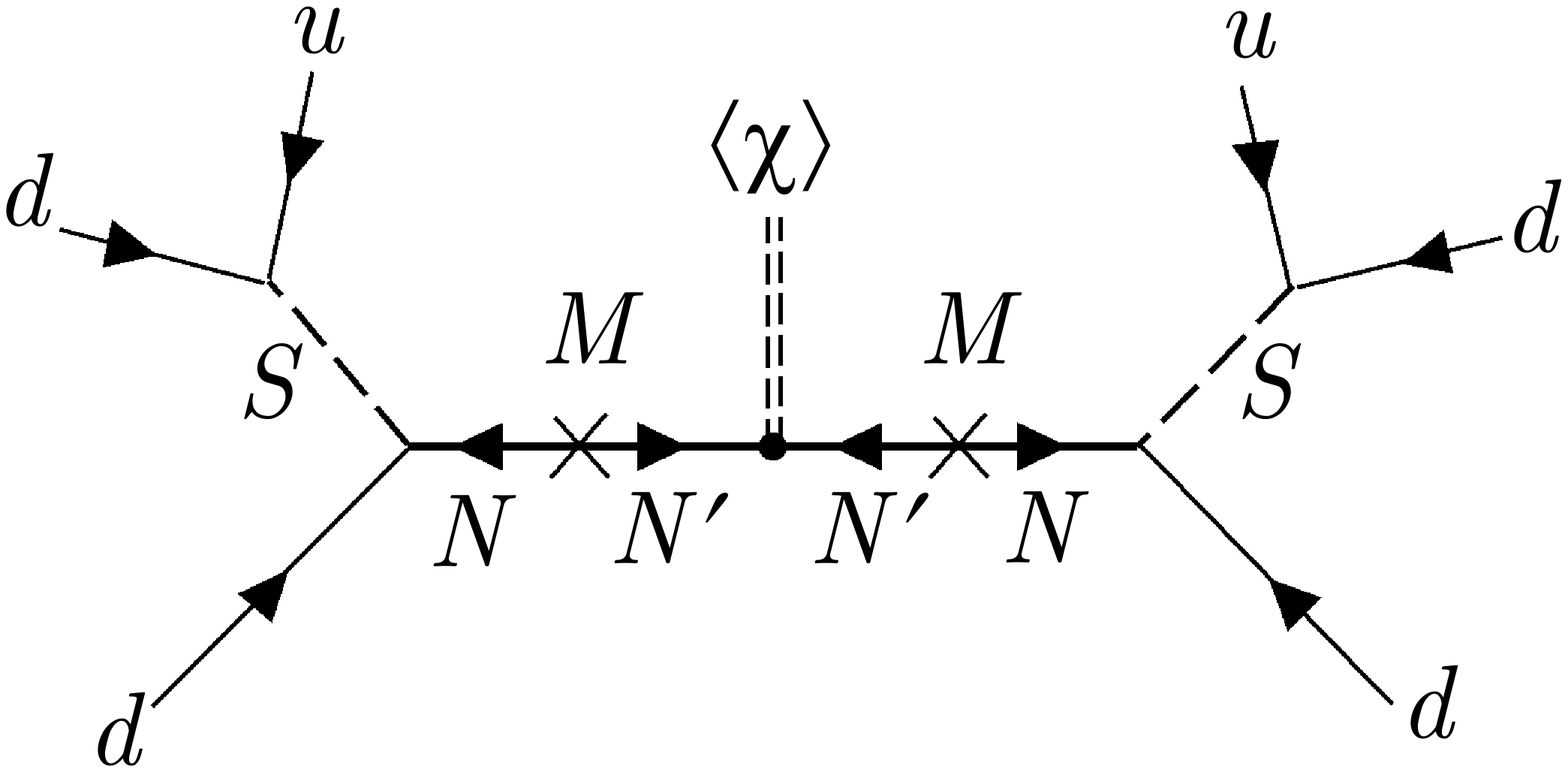} \\
\vspace{-2.4cm}
\includegraphics[angle=270,width=8cm]{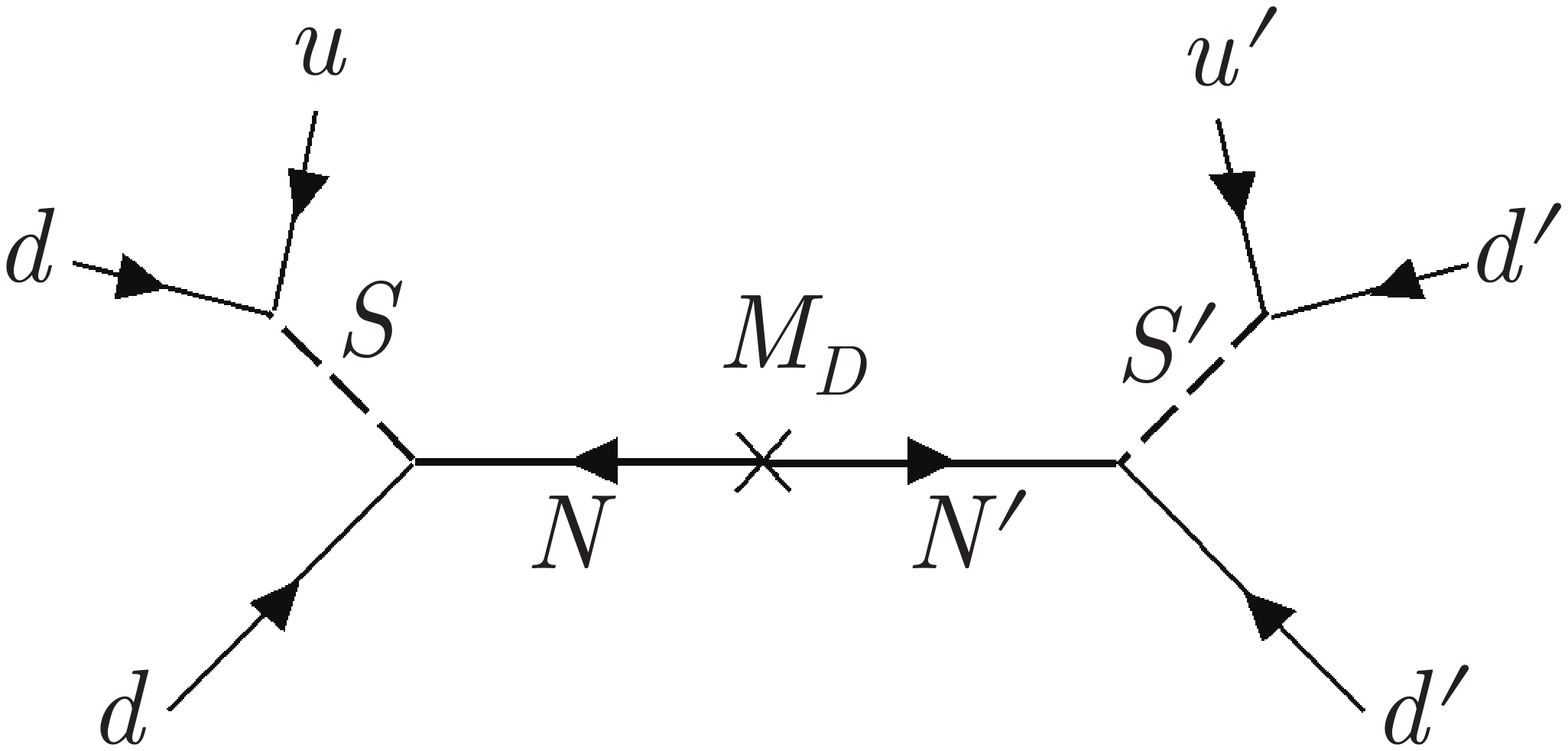}
\vspace{-1.2cm}
\caption{
\label{fig2}
Upper diagram generates  $n - \tilde n$ mixing in low scale baryo-majoron model via exchange of 
heavy Dirac fermion $\cN+\cN'$  when $\cN$ and $\cN'$ get a small Majorana mass 
$\tilde{M}, \tilde{M}' \sim \langle \chi\rangle $. In the presence of  mirror sector containing 
the twin quarks $u',d'$ connected to $\cN'$, lower diagram would generate $n-n'$ mixing 
which conserves the combination of Baryon numbers $B-B'$, without insertion of $\chi$ field. 
}
\end{center}
\end{figure}

One can discuss a simple seesaw-like scenario for generation of terms (\ref{chi}),  
along the lines suggested in ref. \cite{Berezhiani:2005hv,baryo-majoron}. 
%
%The easiest  way  to construct the operators via renormalizable couplings is 
%to use a seesaw-like scenario involving 
%
Let us introduce gauge singlet Weyl fermions, $\cN$ with $Q=-1$ and $\cN'$ with $Q=1$. 
 These two together form a heavy Dirac particle with a large mass $M_D$. 
 Both $\cN$ and $\cN'$ can be coupled to  
scalar $\chi$ ($Q=2$) and get the Majorana mass terms $\sim \langle \chi \rangle =\upsilon_\chi$,
from the VEV of the latter.   
We introduce also a color-triplet scalar $S$, with mass $M_S$ with $Q=-2/3$, 
having precisely the same gauge 
quantum numbers as the right down-quark $d_{(R)}$. 
Consider now the Lagrangian terms 
\be{NNpr}
S u d  +   S^\dagger  d \, \cN + 
M_D \cN \cN' +  \chi^\dagger \cN^2 +  \chi \cN^{\prime 2}  + {\rm h.c.} 
%h_1 S u d  + h_2 S q q  +  h S^\dagger  d \, \cN + 
%M \cN \cN' + \tilde{g} \chi \cN^2 + \tilde{g}' \chi^\dagger \cN^{\prime 2}
\ee
%We assume now that the induced Majorana mass terms of $\cN$ and $\cN'$, 
%$\tilde M = \tilde g f_B$ and $\tilde{M}' = \tilde{g}' f_B$ 
In this way,  diagram shown in Fig. \ref{fig2},
after integrating out the heavy fermions $\cN + \cN'$, induces  $D=10$ operators (\ref{chi}), 
with $M^6 \sim M_D^2 M_S^4$.

 %%%%%%%%%%%%%%%%%%%%%%%%%%%
%%%%%%%%%%%%%%%%%%%%%%%%%%

%\footnote{

Low scale baryon number violation was suggested in Ref. \cite{Berezhiani:2005hv}, 
in a model which was mainly designed for inducing  
neutron -- mirror neutron oscillation $n-n'$. 
 This model treats $\cN$ and  $\cN'$ states  symmetrically: their Majorana masses 
 $\tilde{M}$ and $\tilde{M}'$ are equal, while in addition to couplings (\ref{NNpr}), 
 there are terms that couple $\cN'$ to  $u', d'$ and $S'$ states
 %, partners of $u,d$ and $S$ 
 from hidden mirror sector  with a particle content identical to that of ordinary one 
 (for review, see e.g. \cite{Berezhiani:2003xm}).  
%of ref. \cite{Berezhiani:2005hv}. 
%In addition to couplings (\ref{NNpr}), it includes $h_1 S' u' d'  + h_2 S' q' q'  +  
%h S^{\prime\dagger}  d \, \cN' $ where $u',d'$ and $S'$ are mirror partners  
Hence, the lower diagramm of Fig. \ref{fig2} induces $D=9$ operator $(1/M)^5uddu'd'd'$ 
with $\cM^5 = M_D M_S^4$, and thus $n-n'$ mixing  with 
\be{nn'}
\dm_{nn'} \sim \frac{\Lambda_{\rm QCD}^6}{M_D M_S^4} 
\sim \left(\frac{10~\rm TeV}{\cM} \right)^5 
%\left(\frac{10~\rm TeV}{M_S} \right)^4 
\times 10^{-15}~{\rm eV}
\ee
which corresponds to $n-n'$ oscillation time $\tau_{nn'} \sim 1$~s. 
Hence, in this case  $n-n'$ mixing should be a dominant effect, since two sector share the common $Q=B-L$.  
 between ordinary and mirror particles, 
while $n-\nbar$ mixing which breaks $Q$ is suppressed by the small VEV $\upsilon_\chi$:  
\be{delta-ratio}
\dm_{n\nbar} \leq  \frac{\upsilon_\chi}{M_D} \dm_{nn'}
\ee
Therefore, assuming that $\dm_{nn'} < 10^{-15}$~eV and $M_D > 1$ TeV, for obtaining 
$\dm_{n\nbar} > 10^{-25}$~eV one needs $\upsilon_\chi > 100$~eV. In this case the Galactic 
contribution in $V_n$ becomes irrelevant, but the possibility of having $\lambda < 1$~AU 
remains robust.

Let us remark, that since $n-n'$ mixing conserves $Q=B-L$, baryophotons 
interact symmetrically with ordinary and mirror neutrons, and thus should have 
no effect on $n-n'$ oscillation. 
As a matter of fact, $n-n'$ mixing can indeed be much larger than $n-\nbar$.  
Existing experimental limits on $n-n'$ transition 
%\cite{nnpr-exp} 
allow the neutron$-$mirror neutron oscillation time to be less than the neutron lifetime, 
%which corresponds to $\dm_{nn'} \sim 10^{-15}~{\rm eV}$, 
with interesting implications for astrophysics and particle phenomenology 
\cite{Berezhiani:2005hv,nnpr}. 
 
%\end{document}

\section{Conclusions}

Neutron - antineutron transformation searched with free neutrons can be suppressed 
by the presence of the vector field of baryophotons coupled to $B-L$ charges. 
%that is proportional to the neutron content of the neutral matter. 
Due to assumed baryon number non-conservation these photons should be massive
 with the mass in the range  $10^{-11} - 10^{-21}$ eV. 
 This corresponds to a possible region of $B-L$ potential that is not excluded by  experimental tests 
 of weak equivalence principle (WEP) so that it could suppress the free neutron $n \to \nbar$ transformations. 
 However,  if one learns from nuclear instability search that $n \to \nbar$ transformation exists but it is 
suppressed for free neutrons, then this suppression in principle can be removed by the tuning 
of external magnetic field in the experiment. Weaker $B-L$ fields inducing the potential energy smaller than 
$10^{-16}$ eV, i.e. below the quasi-free condition limit, practically will not be sensed by 
$n \to \nbar$ transformation and therefore cannot be observed in this way. 
STEP experiment for Satellite Test of the Equivalence Principle ~\cite{STEP} proposed some years ago 
claimed the sensitivity of WEP testing to the level $10^{-18}$. 
STEP mission was not pursued. 
Corresponding level of magnitude of $B-L$ potential energy that could be excluded
 in STEP test are also shown in Figure 2. 

Let us remark about the possibility of the kinetic mixing of $B-L$ photons with the regular QED photons. 
Such mixing could make the Equivalence Principle tests potentially different for electrically neutral 
and charged objects, e.g. neutrons and also neutrinos having non-zero $B-L$ could acquire also 
the tiny electric charges. As a matter of fact, the considered $B-L$ potentials can have no effect 
on the oscillations between three neutrinos $\nu_e, \nu_\mu$ and $\nu_\tau$ since their $B-L$ 
charges are equal, but they can be relevant for the active-sterile neutrino (e.g. mirror neutrino) 
oscillations and can suppress them in certain situations.  

Also, $B-L$ charge of the Earth would create a $B-L$ magnetic field due to the Earth rotation. 
%Through kinetic mixing can induce interaction with regular Coulomb charges. 
Question is whether this can lead to any observable effect?

Concluding, if the neutron--antineutron oscillation will be discovered in  free neutron oscillation experiments, 
this will imply limits on $B-L$ photon coupling constant and interaction radius which are considerably 
stronger than present limits form the tests of the equivalence principle. 
The potential $V$ induced by these forces can be excluded down to the values of about $10^{-16}$ eV, 
independently on the interaction radius $\lambda$ of these baryophotons.  
Instead, if $n-\nbar$ oscillation will be discovered via nuclear instability, but not in free neutron oscillations in 
corresponding level, this would indicate towatds the presence of fifth-force mediated by such baryophotons.

\section{Acknowledgments}

Z.B. and Y.K.  thank Arkady Vainshtein for useful discussions. 
The work of A.A. and Z.B. was partially supported by the MIUR 
triennal grant for the Research Projects of National Interest  PRIN 
2012CPPYP7 ``Astroparticle Physics",  
and the work of Y.K. was supported in part by US DOE Grant DE-SC0014558.
This work was reported by Y.K. at the 3rd Workshop ``NNbar at ESS", 
27-28 August 2015, Gothenburg, Sweden. 

\bigskip 

{\bf Note Added:} After this work was completed, it was communicated to us by R.~N.~Mohapatra 
and K.~S.~Babu that they are preparing the work on similar subject.

\end{document}